\documentclass[a4paper, 10pt,titlepage]{article}
\makeatletter                                               
\renewcommand{\@cite}[1]{#1}                                
\makeatother                                                
\newcommand{\onlinecite}[1]{\cite{#1}}                     
\newcommand{\supcite}[1]{{\tiny $^\textrm{\cite{#1}}$  }}

\newcommand{\rcite}[1]{Ref.~\onlinecite{#1}}                

\usepackage{amsmath}
\usepackage{epsfig}
\usepackage[latin1]{inputenc}

\newcommand{\fig}[1]{Fig.~\ref{#1}}
\newcommand{\eq}[1]{Eq.~(\ref{#1})} 
\newcommand{\rapp}[1]{Appendix~\ref{#1}}
\newcommand{\rsec}[1]{Sec.~\ref{#1}}               
\newcommand{\fat}[1]{\mbox{\boldmath$#1$}}
\newcommand{\script}[1]{{\textrm{\scriptsize #1}}}
\newcommand{\su}[1]{_\script{#1}}
\newcommand{\nhat}{\hat{\bf n}}
\newcommand{\xv}{{\bf x}}
\newcommand{\xhat}{\hat{\xv}}
\newcommand{\yv}{{\bf y}}
\newcommand{\yhat}{\hat{\yv}}

\newcommand{\tv}{{\bf t}}
\newcommand{\that}{\hat{\tv}}
\newcommand{\rv}{{\bf r}}
\newcommand{\rhat}{\hat{\rv}}
\newcommand{\nop}{\hspace{-0.1mm}}

\begin{document}
\twocolumn[\large {}]                            
\noindent
{\bf \Large Gyroscope precession in special and general relativity from basic
principles}\\[3mm] 
{\bf Rickard M. Jonsson}\\[2mm] 
{\it Department of Theoretical Physics, Physics and Engineering
Physics, Chalmers University of Technology, and G\"oteborg University, 412
96 Gothenburg, Sweden\\[2mm] 
}
E-mail: rico@fy.chalmers.se\\[2mm]
Submitted: 2004-12-09, Published: 2007-05-01\\
Journal Reference: Am. Journ. Phys. {\bf 75} 463\\
\\
{\bf Abstract}. In special relativity a gyroscope that is suspended in a
torque-free manner will precess as it is moved along a curved path
relative to an inertial frame $S$. We
explain this effect, which is known as Thomas precession, by considering a
real grid that moves along with the gyroscope, 
and that by definition is not rotating 
as observed from its own momentary inertial rest frame. From the basic properties of the 
Lorentz transformation we deduce how the form and rotation 
of the grid (and hence the gyroscope) will evolve relative to $S$. 
As an intermediate step we consider how the grid
would appear if it were not length contracted
along the direction of motion.
We show that the uncontracted
grid obeys a simple law of rotation.
This law simplifies the analysis of spin precession compared to more traditional
approaches based on Fermi transport.
We also consider gyroscope precession relative to an accelerated
reference frame and show that there are
extra precession effects that can be explained in a way
analogous to the Thomas precession. 
Although fully
relativistically correct, the entire analysis is
carried out using three-vectors. 
By using the equivalence principle the formalism can also  be applied to
static spacetimes in general relativity. As an example, we 
calculate the precession of a gyroscope orbiting a static black
hole.

\section{Introduction}\label{intro}
\begin{figure}[h!]
\begin{center}
\epsfig{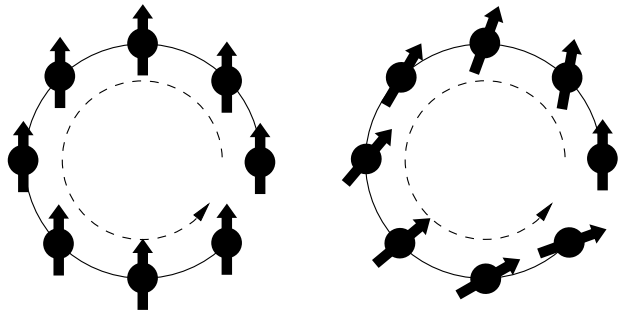}
\caption{A gyroscope transported around a circle. The vectors
correspond to the central axis of the gyroscope at different
times. The Newtonian
version is on the left, the special relativistic version is on the right.} 
\label{fig1}
\end{center} 
\end{figure}

In Newtonian mechanics a spinning gyroscope, suspended 
such that there are no torques acting on it, keeps its
direction fixed relative to an inertial system as we move the gyroscope around a
circle. However, in special relativity
the gyroscope will precess, meaning that the direction of the gyroscope central axis 
will rotate, see \fig{fig1}.

If we denote the circle radius by $R$ and the gyroscope velocity by $v$
we can express the angular velocity $\Omega$ of the gyroscope precession as\supcite{actually}
\begin{equation}
\label{thomas}
\Omega=(\gamma-1)\frac{v}{R}.
\end{equation}
Here $\gamma=1/\sqrt{1-v^2/c^2}$ and $c$ is the speed of
light. Henceforth, unless otherwise stated, we will for
convenience set $c=1$. The precession given by \eq{thomas} is known as
Thomas precession.\supcite{gravitation}
In particular we note that for $v\ll1$ and $\gamma \simeq 1$, the right-hand side of
Eq.~\eqref{thomas} tends to be very small. Thus to obtain a substantial angular velocity
due to this relativistic precession, 
we must have very high velocities (or a very small circular radius).

\begin{figure}[h!]
\begin{center}
\epsfig{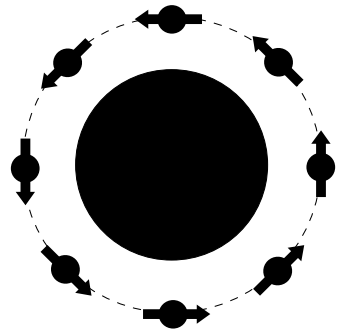}
\caption{A gyroscope transported along a circle at the photon
radius of a static black hole. The gyroscope turns
so that it always points along the direction of motion.} 
\label{fig2}
\end{center} 
\end{figure}

In general relativity the situation becomes even more interesting.
For instance, we may consider a
gyroscope orbiting a static black hole at the photon radius (where
free photons can move in circles).\supcite{photrad}
The gyroscope will precess as depicted in
\fig{fig2} independently of the velocity.

How a gyroscope precesses for these examples can
be derived using four-vectors and Fermi
transport.\supcite{gravitation2} 
Although the Fermi approach is very general, 
it typically results in a set of coupled
differential equations that are rather complicated 
and do not provide much physical insight (see \rapp{Fermi}). 

In the following we will take a different approach. 
We start by discussing why there is Thomas precession in special
relativity. We also derive the exact
relation, \eq{thomas}, using only rudimentary knowledge of special relativity. 
We then consider gyroscope precession with respect to an
accelerated reference frame within special relativity. 
We show that if the gyroscope moves inertially, but the
reference frame accelerates perpendicularly
to the gyroscope direction of motion, the gyroscope will
precess relative to the reference frame.

\begin{figure}[h!]
\begin{center}
\epsfig{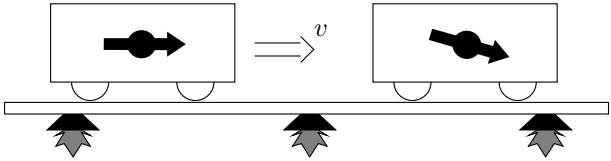}
\caption{A train at two consecutive times moving with
velocity $v$ relative to a platform that accelerates
upward. A gyroscope with a torque free
suspension on the train will precess clockwise for $v>0$.} 
\label{fig3}
\end{center} 
\end{figure}

As an application 
where both the reference frame and the gyroscope accelerates 
we will consider 
a gyroscope on a train that moves along an upward
accelerating platform as shown in \fig{fig3}.

If we neglect the Earth's
rotation, an ordinary platform on Earth behaves just like an accelerated
platform in special relativity (the equivalence principle),\supcite{dinverno}
so the result can be applied to every day scenarios.
The equivalence principle also allows us to apply the analysis to a
gyroscope orbiting a black hole.

Sections~\ref{basis}--\ref{specialconc} assume knowledge of special
relativity. Knowledge of general relativity is assumed in
Secs.~\ref{axi}--\ref{relwork}.

\section{The gyroscope grid}\label{basis}
Although spinning gyroscopes are the typical objects of interest when
discussing relativistic precession effects, we will in the following consider
a grid (say of metal) that we
call the gyroscope grid. 
This grid is by definition not rotating as observed 
from its own momentary inertial rest frame. 
The central axis of an ideal gyroscope 
with a torque-free suspension is, by definition, also non-rotating as observed from its own
momentary inertial rest frame. It follows that the axis of an ideal
gyroscope, which is transported together with the grid, will keep its
direction fixed relative to the grid. Thus, the precession of an
actual gyroscope (assuming it behaves like an ideal gyroscope) 
follows from the behavior of the gyroscope grid. The use of the
grid will also allow us to put the effects of precession due to the
gyroscope grid acceleration on an equal footing with the precession
effects that come from the acceleration of the reference frame.

\subsection{The boost concept}
In special relativity, a Lorentz transformation to a new set
of coordinates, which are non-rotated relative to the original coordinates,
is known as a {\it boost} of the
coordinates.\supcite{boostexact} Equivalently a {\it
physical} boost of an object can be performed.
As seen by an observer at rest in a certain inertial reference  
frame,  a  physical boost of an object by a certain velocity 
is equivalent to performing a boost of the  
observer's reference frame by minus the velocity (while not physically affecting the object).
In particular, boosting an object with respect to the object's
initial rest frame means giving the object the
velocity of the boost and length contracting the object along the
direction of motion.
At times we will use the term {\it pure} boost to stress the
non-rotating aspect of the boost. 
We will also assume that any real 
push of an object (such as the gyroscope grid),
works like a pure boost relative to the momentary inertial rest frame of the object. 

We will next illustrate how three 
consecutive physical boosts of a grid, where the boosts are each
non-rotating as observed from the momentary rest frame of the grid, 
will result in a net rotation. 
The result can be formally derived 
by making successive Lorentz transformations (multiplying matrices),
but from the derivation in the next
section we can also understand how the rotation arises.
The net resulting rotation is the key to the Thomas
precession as presented here.

\subsection{The effect of three boosts}\label{hutt}
\begin{figure}[h!]
\begin{center}
\epsfig{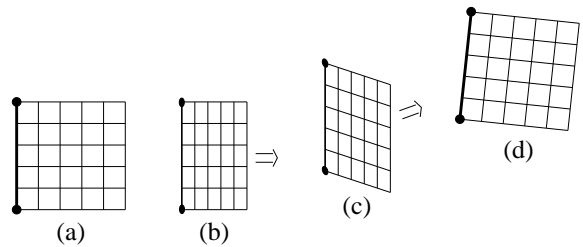}
\caption{
(a) The grid at rest with respect to $S$. 
(b) The grid after a pure boost with velocity $v$ to the
right, relative to $S$. Note the length contraction. (c) The grid after a pure upward boost
relative to a system $S'$ that moves with velocity $v$ to the
right. 
(d) The grid after a pure boost that stops the grid relative to
$S$.}
\label{fig4}
\end{center} 
\end{figure}

Consider a grid in two dimensions,
initially at rest and non-rotating relative to an inertial system $S$. 
We then perform a series of 
boosts of the grid as sketched in \fig{fig4}.
Note especially what happens to the thick bar and its end
points. 

After the first boost by a velocity $v$ to the right, the grid is at rest relative to another inertial
system $S'$. The grid is then given a pure
upward boost relative to $S'$ by a velocity $\delta u'$. Relative
to the original system $S$, the grid will then move in a direction
$\nhat$ up and to the right. 
Through the upward boost the originally vertical grid bars remain 
vertical relative to $S'$; thus they will also remain vertical relative to
$S$, as follows from the Lorentz transformation. 
However, the originally horizontal bars will become rotated.
To understand this rotation, 
consider all of the events along a horizontal bar just as the bar
starts moving upward relative to $S'$. These events are all simultaneous
relative to
$S'$, but relative to $S$ the 
rearmost event (the leftmost event)
will happen first (relativity of simultaneity).
Thus the leftmost part of the bar will have a head start (upward) relative to
the rightmost part, and the bar will therefore become rotated.
Finally, we stop the grid, in other words we make a pure boost in the
$-\nhat$ direction, so that the grid stops relative
to $S$. The effect will be to remove the length contraction in the $\nhat$
direction, that is, to stretch the grid
in the $\nhat$ direction. Through this
stretching we understand that the originally vertical grid bars will
rotate clockwise. Because none of the boosts deform the grid as
observed in the grid's own momentary rest frame, it follows that the
entire final grid will be rotated clockwise relative
to the original grid. 

\subsection{Calculating the precise turning angle}\label{precise}
The upward boost by a velocity $\delta u'$ relative to $S'$
yields an upward velocity $\delta u=\delta u'/\gamma$ (time dilation) as observed from $S$.
Consider now two points separated by a distance $L_{0}$ 
along an originally vertical bar of
the grid, as measured in the grid's own frame. 
As observed in $S'$, the distance between the points after the upward boost
is, due to length contraction, given by $L=L_{0}/\gamma(\delta u')$. 
This distance is also the distance between
the points as observed in $S$, as follows from the Lorentz transformation.
Also, the velocity of the points after the upward boost is $v \xhat + \delta u \yhat$ 
as observed in $S$. 

\begin{figure}[h!]
\begin{center}
\epsfig{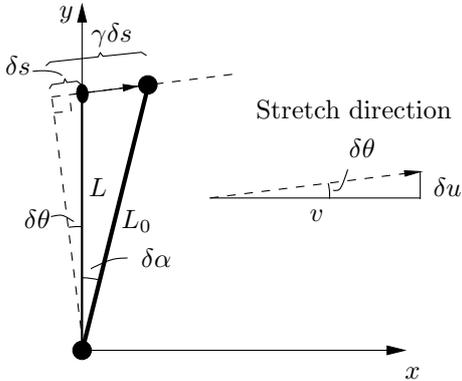}
\caption{The stretch-induced tilt of 
the two points (the filled circles)
due to the final stopping of the grid. 
The distance between the points prior to the stretching, as measured
in the direction of motion, is denoted by $\delta s$.} 
\label{fig5}
\end{center} 
\end{figure}

When we stop the grid, the length expansion (that is, the removal of length
contraction) will shift the topmost point relative to the
lowest point, resulting in a rotation by an angle $\delta \alpha$ 
as depicted in \fig{fig5}. From the definitions in \fig{fig5} it follows that

\begin{equation}\label{hipp}
\tan \delta \theta = \frac{\delta u}{v}, \quad
\sin(\delta \alpha + \delta \theta)=\frac{\gamma \delta
 s}{L_{0}},\quad
\sin \delta \theta=\frac{\delta s}{L}.
\end{equation}
From now on we will assume that $\delta u$ is infinitesimal. 
Because the $\gamma$-factor entering the relation between $L^{}$ and
$L_{0}$ depends on $(\delta u)^2$, we have $L^{}=L_{0}$ to first
order in $\delta u$.
It follows that, to first order in $\delta u$, Eq.~(\ref{hipp}) reduces to 
\begin{equation}\label{hipp2}
\delta \theta= \frac{\delta u}{v}, \quad
\delta \alpha +\delta \theta=\frac{\gamma \delta s}{L_{0}}, \quad
\delta \theta=\frac{\delta s}{L_{0}}.
\end{equation}
From Eq.~(\ref{hipp2}) we find
\begin{equation}\label{ahhh}
\delta \alpha = \frac{\delta u}{v} (\gamma-1).
\end{equation}
Here $\delta \alpha$
is the resultant clockwise angle of rotation of the
grid, after the three consecutive boosts.
The result also applies to a grid that was initially rotated by a
certain angle relative to the grid we considered above. 
To see this, suppose that we perform the three
boosts simultaneously on the two grids. Because the boosts are all
non-rotating as observed in the momentary rest frame of the grids, the
relative angle between the grids must be preserved. Thus \eq{ahhh} gives the
angle of rotation resulting from the three boosts in question, regardless of
the initial rotation of the grid.

For an infinitesimal boost 
in a general direction relative to $S'$,
only the upward directed part of the boost contributes to the rotation.\supcite{note} 
Thus \eq{ahhh} holds
also for this case if we interpret $\delta u$ in \eq{ahhh} as the
part of the infinitesimal velocity change that is perpendicular to the direction of motion.

\subsection{The uncontracted grid}\label{stopcom}
\begin{figure}[h!]
\begin{center}
\epsfig{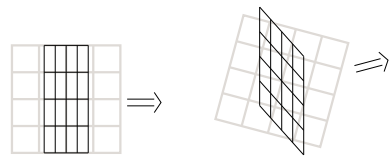}
\caption{The real grid (black thin lines) and the corresponding
imagined uncontracted grid
(grey thick lines) before and after a large upward boost.}
\label{fig6}
\end{center} 
\end{figure}

For a grid in motion relative to a certain specified reference frame
(for example an inertial frame), we now introduce
what we call the {\it uncontracted} grid. 
This grid is obtained by imagining the real grid without
length contraction along the direction of motion. The idea is illustrated in \fig{fig6}.

From \fig{fig4}, we note that the imagined uncontracted grid,
immediately before and after the infinitesimal
upward boost, is identical (in form and rotation) to the initial and the final
actual grid respectively.
From the discussion at the end of 
Sec.~II~C it therefore follows that 
for any real grid moving on a plane that receives an infinitesimal boost 
(non-rotating as observed in the grid's own momentary inertial rest frame) by a
velocity $\delta u$ perpendicular to the direction of motion, the
corresponding uncontracted grid will rotate an angle given by \eq{ahhh} as
\begin{equation}
\label{ahhh2}
\delta \alpha = \frac{\delta u}{v} (\gamma-1).
\end{equation}
Henceforth we will always describe the
gyroscope grid in terms of the 
uncontracted grid. If we have found the evolution of the uncontracted grid for a particular
path, we can always find the observed 
real grid by length contracting the uncontracted grid in the momentary direction of motion.

\subsection{Circular motion}
\begin{figure}[h!]
\begin{center}
\epsfig{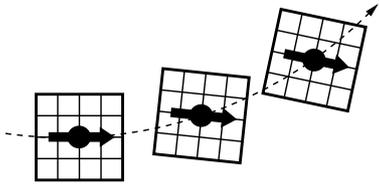}
\caption{A gyroscope grid at successive time
steps.
Both the grid and the gyroscope are depicted as they would
be observed if they were uncontracted.} 
\label{fig7}
\end{center} 
\end{figure}

Consider a gyroscope grid moving with velocity $v$ along a circle of radius $R$. 
During a time step $\delta t$, the grid receives an infinitesimal boost
perpendicular to the direction of motion. In the inertial frame of the
circle the perpendicular velocity change is given by
\begin{equation}\label{addeq}
\delta u=\frac{v^2}{R} \delta t.
\end{equation}
The corresponding uncontracted grid will rotate an
angle according to \eq{ahhh2} during the boost.
At the next time step 
there is a new boost and a new induced rotation.
It follows that there is an ongoing precession of
the uncontracted grid as depicted in \fig{fig7}.
The angular velocity of rotation is given by Eqs.~\eqref{ahhh2} and
\eqref{addeq} as 
\begin{equation}
\label{this1}
\frac{\delta \alpha}{\delta t}=(\gamma-1) \frac{v}{R}.
\end{equation}
Thus we have derived the Thomas precession given by Eq.~\eqref{thomas}.
Note that Eq.~\eqref{this1} describes how fast the imagined uncontracted grid rotates. 

\subsection{The mathematical advantage of the uncontracted grid}
We have shown that the uncontracted grid evolves according to a simple law of
rotation. The central axis of a gyroscope, if it
were not length contracted along the direction of motion, obeys the same
simple law of rotation. The actual axis of a gyroscope, however,
changes its length over time, and its angular velocity would not be as
simple as that given by \eq{this1}.
A differential equation for the evolution of the actual axis, 
would hide the simple dynamics of a rotation and a superimposed length
contraction, which would complicate the analysis.
Similarly, the standard approach to calculating gyroscope precession, 
which uses the Fermi transport
equation for the spin vector of the gyroscope, is also comparatively
complicated (see \rapp{Fermi} and \rcite{rickardfermi}
for further details).

\subsection{Comments on the uncontracted grid}\label{stopcom2}
Although we may think of the uncontracted grid as a mathematically convenient
intermediate step in finding the actual grid, there is more to this concept.
As follows from its definition, the uncontracted grid corresponds directly to 
the grid as experienced in a system that moves with the grid and that is related to the
reference frame in question by a pure boost. 

Consider a special relativistic scenario of a gyroscope grid
suspended in a torque free manner inside a satellite. The satellite
uses its jet engines to move along a smooth simple closed curve on a plane.
We want to measure from the
satellite the precession angle of the
gyroscope grid after a full orbit. 
If we assume that there are a couple of suitably placed fixed stars, we can
use their direction as
observed from the satellite at the initial and final point of the
orbit (which coincide), as guidelines to establish a reference
system within the satellite.
For this scenario the uncontracted grid 
is the physical object in which we are interested, because it's
orientation precisely corresponds to the orientation of the actual gyroscope grid relative to the star
calibrated reference frame of the satellite. 
In particular, if the uncontracted grid has rotated a certain
angle after the full orbit, so has the actual gyroscope grid as measured 
from the satellite.

\section{Boosting the reference frame}
\begin{figure}[h!]
\begin{center}
\epsfig{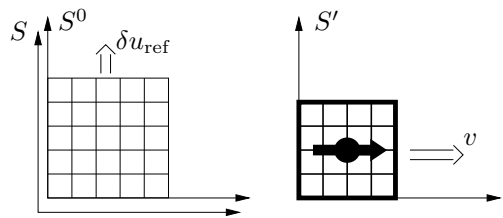}
\caption{Boost of the reference frame (of which the
depicted thin grid is a small part) upward by a velocity
$\delta u_{\script{ref}}$. The velocity of the
gyroscope grid is maintained.}
\label{fig8}
\end{center} 
\end{figure}

Now let us consider the effect of a boost of the reference frame rather than
of the gyroscope grid. To make the analogy with the discussion
in \rsec{basis} clearer, we consider a real grid as a reference frame. 
We assume that the reference frame initially is at rest relative to an
inertial system $S$, and is then boosted upward
so that it is at rest with respect to another inertial system $S^0$,
see \fig{fig8}.
The gyroscope grid is assumed to move with constant speed $v$ to the right
as observed in $S$.

\begin{figure}[h!]
\begin{center}
\epsfig{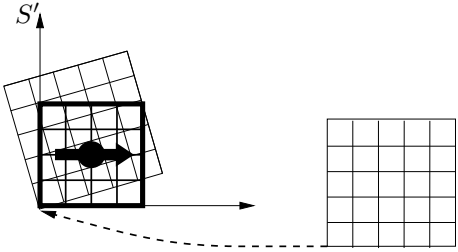}
\caption{Relative to the gyroscope system, the reference
frame (thin lines), of which we illustrate a certain
part, rotates during the boost. The
reference frame is depicted as it would appear if it was not
length contracted relative to the gyroscope system.}
\label{fig9}
\end{center} 
\end{figure}

Relative to the gyroscope system, the reference frame initially
moves to the left, and is then (due to the boost) given an upward velocity $\delta
u_{\rm ref}/\gamma$ (time dilation). Because the reference frame moves
relative to the gyroscope system, the reference frame is length
contracted along the direction of motion.
However, we can imagine the reference frame
without the length contraction.
Analogous to the discussion in \rsec{basis},
the uncontracted reference frame will rotate during the
boost, as depicted in \fig{fig9}. 

The net counterclockwise angle of rotation for the reference frame is found by substituting
$\delta u$ by $\delta u_{\rm ref}/\gamma$ into \eq{ahhh2}:
\begin{equation}
\label{ai}
\delta \alpha= \frac{\delta u_{\script{ref}}}{\gamma} \frac{\gamma-1}{v}.
\end{equation}
Note that removing the length contraction of the uncontracted reference frame
relative to the gyroscope grid yields the same relative configuration
as removing the length contraction of the gyroscope grid relative to the reference frame. 
It follows that the upward boost of the reference frame yields a
clockwise rotation of the uncontracted gyroscope grid, described by
\eq{ai}, relative to the reference frame. 
This relative rotation
is precisely the rotation in which we are 
interested. Note in particular that an upward boost of the
reference frame yields a clockwise relative rotation just as an upward boost
of the gyroscope grid yields a clockwise relative rotation.

\section{Three dimensions}
In the two-dimensional reasoning of Secs.~II and III, the
induced rotation
occurred in a plane spanned by the velocity vector and the
vector for the velocity change. 
For more general three-dimensional motion
and velocity changes, the induced rotation should still occur in a
plane spanned by these two vectors. 
The axis of rotation can therefore be
expressed in terms of the cross product of these two vectors. 
Let us introduce $\delta \fat{\alpha}$ as a vector whose direction
indicates the axis of rotation and whose magnitude corresponds to the angle of
rotation for the uncontracted gyroscope grid relative to the
reference frame. Also, let ${\bf v}$ be the velocity vector of the
gyroscope relative to the reference frame. The three-vector analog
of \eq{ahhh2} for a velocity change $\delta {\bf u}_\script{gyro}$ of
the gyroscope, using the identity $(\gamma-1)=\gamma^2 v^2/(\gamma+1)$, 
can be written as 
\begin{equation}\label{rr1}
\delta \fat{\alpha}=\frac{\gamma^2}{\gamma+1} ({\delta {\bf u}_\script{gyro}} \times 
{{\bf v}}).
\end{equation}
The corresponding vector analog of \eq{ai} for a velocity change
$\delta {\bf u}_\script{ref}$ of the reference frame
is given by
\begin{equation}\label{rr2}
\delta \fat{\alpha}=\frac{\gamma}{\gamma+1} ({\delta {\bf u}_{\script{ref}}} \times 
{{\bf v}}).
\end{equation}
Note that the cross product selects only the
part of the velocity change that is perpendicular to the relative
direction of motion.

Consider now an infinitesimal boost of both the gyroscope
and of the reference frame. 
If we assume that we start by boosting the gyroscope, which gives a
velocity change $\delta {\bf u}_{\rm gyro}$, 
this boost yields a rotation according to
\eq{rr1}. Subsequently boosting the reference frame by a velocity
$\delta {\bf u}_{\rm ref}$ yields another
rotation given by \eq{rr2}, but ${\bf v}$ should be replaced by ${\bf v}+\delta {\bf u}_{\rm gyro}$. 
However, to first order in $\delta {\bf u}_{\rm ref}$ and $\delta {\bf u}_{\rm
 gyro}$ 
this replacement
does not affect \eq{rr2}.
Because infinitesimal rotations can be added (to first
order in the magnitude of the rotations), 
it then follows that the net
rotation is given by Eqs.~\eqref{rr1} and \eqref{rr2} as
\begin{equation}\label{bx}
\delta \fat{\alpha}=\frac{\gamma^2}{\gamma+1} ({\delta {\bf
u}_\script{gyro}} \times {{\bf v}})+\frac{\gamma}{\gamma+1} ({\delta {\bf
u}_{\script{ref}}} \times {{\bf v}}).
\end{equation}
Now consider a continuously accelerating reference frame and gyroscope
grid. Relative to an inertial system in which the reference
frame is momentarily at rest, 
we have $\delta {\bf u}_\script{gyro}={\bf a}_{\rm gyro}
\delta t$ and $\delta {\bf u}_{\rm ref}={\bf a}_{\rm ref}
\delta t$ for a time step $\delta t$. 
We substitute these relations into \eq{bx} and obtain the net angular 
velocity vector $\fat{\Omega}=\delta \fat{\alpha}/\delta t$ for the gyroscope grid 
rotation relative to the
reference frame as
\begin{equation}
\label{yes}
\fat{\Omega}=\frac{\gamma}{\gamma+1}[ \gamma {\bf
a}_{\rm gyro} + {\bf a}_{\rm ref}
] \times {\bf v}.
\end{equation}
Because we are interested in how the gyroscope grid rotates relative to
the accelerating reference frame, it can be useful to express the motion
relative to the reference frame. Consider 
a path
with local curvature radius $R$ and curvature direction $\nhat$, fixed to the reference frame. 
In \rapp{curveref} we show that for motion along this path we have
(just like in Newtonian mechanics)
\begin{equation}\label{yes2}
[{\bf a}_{\rm gyro}]_\perp=[{\bf a}_{\rm ref}]_\perp + v^2 \frac{\nhat}{R}.
\end{equation}
Here ${\bf a}_{\rm gyro}$ and ${\bf a}_{\rm ref}$ refer to
the accelerations relative to an inertial frame in which the reference frame is momentarily at rest.
The notation $\perp$ denotes the part of the acceleration that is perpendicular to the
direction of motion for the gyroscope grid. Because of the cross product in \eq{yes}, 
the perpendicular part
of the acceleration is the only part that matters for $\fat{\Omega}$. 
If we substitute ${\bf a}_{\script{gyro}}$ from
\eq{yes2} into \eq{yes} and simplify the resultant expression, we find
\begin{equation}\label{yes3b}
\fat{\Omega}=(\gamma -1) \Big(\frac{\nhat}{R} \times {\bf v}
\Big) + \gamma(
{\bf a}_{\rm ref}\times {\bf v}).
\end{equation}
The first term on the right-hand side of Eq.~\eqref{yes3b} has the same form as the standard Thomas
precession term given by \eq{thomas}. The second term corresponds to
both direct effects of
rotation from the reference frame acceleration and to the indirect effects of this
acceleration, because the acceleration of the
gyroscope grid relative to an inertial frame depends on the reference
frame acceleration in this formulation.
Equation~\eqref{yes3b} matches the formally derived
Eq.~(51) of \rcite{rickardfermi}.

\section{Applications }\label{applications}
In this section we discuss applications of the derived formalism,
Eq.~\eqref{yes3b}, for gyroscope
precession relative to an accelerating reference frame.

\subsection{Motion along a horizontal line}
\begin{figure}[h!]
\begin{center}
\epsfig{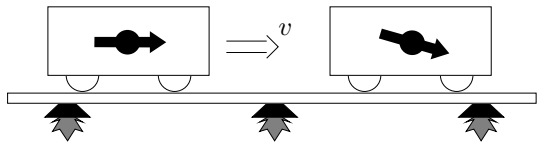}
\caption{A train with a gyroscope moving relative to
an accelerating platform observed at two successive times.} 
\label{fig10}
\end{center} 
\end{figure}

Consider a special relativistic scenario of a train moving along a
horizontal line relative to a platform that continually accelerates
upward relative to an inertial frame (see \fig{fig10}). 

On the train a gyroscope is suspended so that there are no torques acting on
it as observed from the train. For the special case of motion along a straight line 
relative to the
reference frame (the platform in this case), we have $1/R=0$.
If we define ${\bf g}=-{\bf a}_{\script{ref}}$ as the local acceleration of an
object dropped relative to the platform, \eq{yes3b} reduces to
\begin{equation}\label{yes3bb}
\fat{\Omega} = \gamma ({\bf v} \times {\bf g}).
\end{equation}
Note that both the gyroscope and the platform
reference frame accelerate with respect to an inertial frame, and
hence we expect two precession effects. Both of these effects are included
in the single term on the right-hand side of \eq{yes3bb}. 
Let $\Omega$ denote the clockwise precession rate, $v$ the
velocity to the right, and $g$ the the downward acceleration of
dropped object relative to the platform. Then Eq.~\eqref{yes3bb} gives
\begin{equation}\label{perfa2}
\Omega=\gamma vg.
\end{equation}
Note that the uncontracted grid, whose rotation with respect to the
platform reference frame is given by \eq{perfa2}, corresponds to the
grid as experienced by an observer on the train 
(as discussed in Sec.~II~G). 
Thus we obtain the angular velocity relative to
the train by multiplying the right-hand side of \eq{perfa2} by
$\gamma$ to account for time dilation.\supcite{noteg}
Relative to the train the gyroscope thus precesses at a steady
rate $\Omega_0$ (clockwise as depicted) given by
\begin{equation}
\label{matcha}
\Omega_0
=\gamma^2 vg.
\end{equation}
If we assume the train velocity to be low and introduce the proper factor of $c^2$ to enable us
to express $g$ and $v$ in SI units, we have $\Omega_0 \approx
vg/c^2$. For a train with a velocity of 50\,m/s and a platform acceleration
corresponding to that of a dropped apple on the Earth, we obtain 
\begin{equation}\label{matcha2}
\Omega_0 \approx\frac{50 \frac{\textrm{m}}{\textrm{s}} \times 9.81
\frac{\textrm{m}}{{\textrm{s}}^2}}{(3 \times 10^8 \frac{\textrm{m}}{\textrm{s}})^2}\approx 5 \times 10^{-15} \textrm{rad}/\textrm{s}.
\end{equation}
This special relativistic
scenario mimics a train moving along a straight platform on the
Earth (neglecting the Earth's rotation). It follows that
precession effects due to gravity are small for everyday scenarios on
the Earth.

\begin{figure}[h!]
\begin{center}
\epsfig{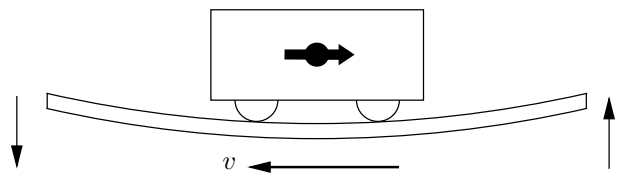}
\caption{A sketch of the rail and the train observed from an inertial system
where the train is momentarily at rest. } 
\label{fig11}
\end{center} 
\end{figure}

Because a torque-free gyroscope precesses relative to the train,
it follows that the train  has a proper rotation, meaning
that the train rotates as observed from its own momentary inertial rest
frame. This rotation 
can be understood without reference to the gyroscope
precession. The heart of the matter lies (as is often the case) in simultaneity. 
Let $S$ be an inertial system where the rail of the continuously
accelerating platform is at
rest momentarily (at $t=0$). As observed in $S$, the
horizontal straight rail will first move downward (when $t<0$),
decelerate to be at rest at $t=0$, and then accelerate
upward. Consider now all the events along a section of the rail, 
when the rail is at rest in $S$. 
Relative to the train's momentary inertial rest frame $S'$, which moves with
velocity $v$ to the right as observed from $S$, the rightmost of the
events along the rail will occur first. Thus relative to $S'$, when the rail at the rear
end of the train has no vertical motion, the rail at the front end (and thus
the train's front end) will already have an upward velocity. 
Hence a train moving as depicted in \fig{fig10} has a proper counterclockwise rotation. 
By this reasoning we can verify the
validity of \eq{matcha}.
We also understand that as observed from $S'$, the rail
is not straight but is curved as depicted in \fig{fig11}.

\subsection{Following the geodesic photon}\label{geophot}
\begin{figure}[h!]
\begin{center}
\epsfig{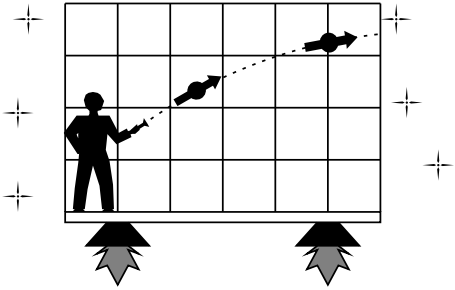}
\caption{A free photon will in general follow a curved path relative to an
accelerated reference frame. A gyroscope transported along
such a path will keep pointing along the path if it did so initially.} 
\label{fig12}
\end{center} 
\end{figure}

As another application we now study the precession of
a gyroscope that follows the spatial trajectory of a free photon.\supcite{marek} 
From \eq{hepp} it follows that the trajectory of a free photon (set $v=1$
and ${\bf a}_{\script{particle}}=0$) as
observed relative to an accelerating 
reference frame satisfies
\begin{equation}\label{hejt}
\frac{\nhat}{R}={\bf g}_\perp.
\end{equation}
Here ${\bf g}_\perp$ is the part of ${\bf g}$ that is perpendicular to
${\bf v}$ (recall that ${\bf g}=-{\bf a}_{\script{ref}}$). 
If we substitute the curvature given by \eq{hejt} into \eq{yes3b}, we
find that the gyroscope grid angular velocity is given by
\begin{equation}\label{phottraj0}
\fat{\Omega}={\bf v} \times {\bf g}_\perp.
\end{equation}
Consider now a normalized vector $\that$ directed along the spatial
direction of motion. The time derivative of $\that$ relative
to the reference frame satisfies $d\that/dt=v \nhat/R$. It follows that $\that$ rotates with an
angular velocity 
$\fat{\Omega}\su{$\bf \hat{t}$}={\bf v} \times \nhat/R$. 
If we substitute the curvature $\nhat/R ={\bf g}_\perp$ given by
\eq{hejt}, into this expression for $\fat{\Omega}\su{$\bf \hat{t}$}$, we obtain 
\begin{equation}
\label{this}
\fat{\Omega}\su{$\bf \hat{t}$}={\bf v} \times {\bf g}_\perp.
\end{equation}
If we compare Eqs.~\eqref{this} and \eqref{phottraj0}, we see that $\that$ rotates 
with the same angular velocity as the gyroscope grid.
It follows that a gyroscope transported along a spatial
trajectory of a free photon will keep pointing along the direction of
motion if it did initially (see \fig{fig12}).

If we imagine a static reference frame outside the event
horizon of a static black hole, then locally this reference frame
behaves just like an accelerated reference frame in special relativity
(the equivalence principle). 
Hence a gyroscope outside of a black hole that follows the path of a
free photon, such as a circle at
the photon radius, will not precess
relative to the forward direction of motion, as depicted in
\fig{fig2} in the introduction. 

\section{Conclusions}\label{specialconc}
We have seen how the basic principles of special relativity can be used to
derive a simple but exact three-vector formalism of spin
precession with respect to an accelerating reference frame. The
precession is given by \eq{yes3b} as
\begin{equation}\label{yes3bconc}
\fat{\Omega}=(\gamma -1) \Big(\frac{\nhat}{R} \times {\bf v}
\Big) + \gamma({\bf a}_{\rm ref}\times {\bf v}),
\end{equation}
where $\nhat/R$ is the curvature of the gyroscope path relative to the accelerated reference
frame. Recall that $\fat{\Omega}$ describes the rotation of a gyroscope axis
as we imagine it without length contraction along the direction of
motion. In the following, knowledge of general relativity is assumed.

\section{Axisymmetric spatial geometries and effective rotation vectors}\label{axi}
\begin{figure}[h!]
\begin{center}
\epsfig{figure=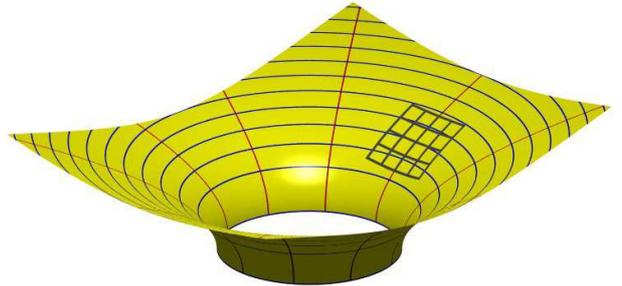,width=8cm}
\caption{
Sketch of the spatial geometry of a symmetry plane outside a black
hole. The local static reference frame shown (the square grid) has a proper
acceleration outward. For a sufficiently small such reference frame 
it works just like an accelerated reference frame in special relativity.}
\label{fig13}
\end{center} 
\end{figure}

In a static spacetime such as that of a Schwarzschild black hole, the global
static reference frame locally corresponds to the accelerated reference
frames we have considered in special relativity. 
If we integrate the infinitesimal rotations from $\fat{\Omega}$ given
by either Eqs.~\eqref{yes} or \eqref{yes3b}, 
we can find the net rotation of a gyroscope that is transported along
a given spatial path.
Note, however, that $\fat{\Omega}$
describes how the gyroscope grid rotates relative to a frame
that is parallel
transported with respect to the local spatial geometry associated with the
reference frame. Thus directly integrating the effects of
rotation from $\fat{\Omega}$ gives the rotation relative to a frame
that is parallel transported with respect to the global spatial
geometry. In \fig{fig13} we illustrate a section of the spatial geometry of
an equatorial plane of a static black hole.

Suppose then that we consider
motion in the equatorial plane of some axisymmetric geometry. 
For instance, we might be interested in the the net rotation of a
gyroscope (grid) after a closed orbit around the center of symmetry (not necessarily a circular orbit). 
We need then 
take into consideration
that a parallel transported frame will be rotated relative
to its initial configuration after a complete orbit due to the
spatial geometry. 
To deal with this complication 
we introduce a new reference frame that rotates relative to the local
coordinates, spanned by the polar vectors $\rhat$ and $\fat{\hat{\varphi}}$, in the
same manner as a parallel transported reference frame does on a plane.
In other words, if we
consider a counterclockwise displacement ($\delta\varphi$,$\delta r$), then relative to the local
vectors $\rhat$ and $\fat{\hat{\varphi}}$, the new reference
frame should rotate $\delta \varphi$ clockwise. Such a
``would-be-flat'' reference frame always returns to its initial configuration after
a full (closed) orbit.

The line element for a two-dimensional axisymmetric spatial geometry
can be written in the form
\begin{equation}\label{burt}
ds^2=g_{rr} dr^2 + r^2 d\varphi^2.
\end{equation}
With respect to such a geometry it is easy to show that the angular velocity of a parallel
transported frame relative to a ``would-be-flat'' frame is given by\supcite{rickardfermi}
\begin{equation}\label{katt}
\fat{\omega}\su{space}=\frac{1}{r}\Big(\frac{1}{\sqrt{g_{rr}}}-1 \Big) {\bf
v} \times \rhat.
\end{equation}
Because infinitesimal rotation vectors can be added (to lowest
order), it follows from \eq{katt} and \eq{yes3b} that the gyroscope grid rotation relative to the
would-be-flat frame is given by
\begin{equation}\label{katt2}
\fat{\Omega}_{\rm eff}=
(\gamma-1)\Big(\frac{\nhat}R \times {\bf v} \Big) 
-\gamma ({\bf g} \times {\bf v})
+\frac{1}{r} \Big(\frac{1}{\sqrt{g_{rr}}}-1
\Big) {\bf v} \times \rhat.
\end{equation}
Alternatively we could express $\fat{\Omega}_{\rm eff}$
in terms of the gyroscope acceleration ${\bf a}\su{gyro}$
relative to a local freely falling (inertial) 
frame momentarily at
rest relative to the static reference frame. 
If we use \eq{yes}
and add the rotation due to the spatial geometry as described by
\eq{katt}, we find
\begin{equation}
\label{katt2b}
\fat{\Omega}_{\script{eff}}=
\frac{\gamma^2}{\gamma+1} ({\bf a}\su{gyro}\times {\bf v}) -
\frac{\gamma}{\gamma+1}({\bf g} \times {\bf v}) 
+\frac{1}{r} \Big(\frac{1}{\sqrt{g_{rr}}}-1
\Big) {\bf v} \times \rhat.
\end{equation}
Note that the time $t$ implicitly entering in Eqs.~\eqref{katt2} and \eqref{katt2b}
through $\Omega\su{eff}=d \alpha/dt$ is the local proper time
for a static observer. We obtain the net induced rotation of a
gyroscope in closed orbit by integrating the effects of the
infinitesimal rotations given by either Eq.~\eqref{katt2} or Eq.~\eqref{katt2b}. 

\section{Circular orbits in static sphe\-rically symmetric spacetimes}\label{circ}
For circular motion in a spatial symmetry plane of a static spherically symmetric spacetime, the
direction of the rotation vector $\fat{\Omega}_{\script{eff}}$ is constant
(directed perpendicularly to the plane of motion) in the coordinate basis of the
would-be-flat reference frame. 
For a counterclockwise motion
the clockwise angular velocity of precession is then (with
$\nhat =- \rhat$ and ${\bf g}=-g \rhat$)
given by \eq{katt2} as 
\begin{equation}\label{hola0}
\Omega_{\rm eff}=(\gamma-1)\frac{v}{R} - \gamma g v +
\frac{v}{r} \Big(\frac{1}{\sqrt{g_{rr}}}-1 \Big).
\end{equation}
It is easy to show that the curvature radius of a circle at a certain
$r$, for a geometry of the form of \eq{burt}, is given by $R=r
\sqrt{g_{rr}}$. If we substitute this result into \eq{hola0}, we find
\begin{equation}\label{hola}
\Omega_{\rm eff}=\frac{v}{r}
\Big(\frac{\gamma}{\sqrt{g_{rr}}}-1 \Big) - \gamma g v.
\end{equation}
For a general spherically symmetric static spacetime, the line element of a
radial line can be written in the form
\begin{equation}\label{br}
d\tau^2=g_{tt}(r) dt^2 - g_{rr}(r) dr^2.
\end{equation}
Note that $g_{rr}$ is positive as defined here (to match the
definition in \rsec{axi}).
From \eq{br} it is easy 
to derive the local
acceleration of a freely falling particle momentarily at rest. The result is
\begin{equation}\label{gfall}
g=\frac{1}{2 g_{tt} \sqrt{g_{rr}} } \frac{\partial g_{tt}}{\partial r}.
\end{equation}
So here we have an explicit expression for the $g$ which enters the
expression for $\Omega_{\script{eff}}$ in \eq{hola}. We are now ready to consider a
specific example.

\subsection{The Schwarzschild black hole}
For a Schwarzschild black hole (using standard
coordinates and $c=G=1$) we have
\begin{subequations}
\label{tro}
\begin{align}
g_{tt}&= (1-2M/r)\\
g_{rr}&=(1-2M/r)^{-1} \label{tro.rr}. 
\end{align}
\end{subequations}
We substitute these two expressions into \eq{gfall} and find
\begin{equation}\label{gref}
g=\frac{M}{r^2 \sqrt{1-2M/r}}.
\end{equation}
If we use \eq{gref} and \eq{tro.rr} in \eq{hola}, we obtain
\begin{equation}\label{hatt}
\Omega_{\rm eff}=\frac{\gamma v}{r \sqrt{1-2M/r}}
\big( 1 - 3M/r \big) - \frac{v}{r}.
\end{equation}
Equation~\eqref{hatt} gives the precession rate as a function of $r$ and $v$. 
For constant velocity $v$ we obtain the net rotation
after a full orbit by multiplying the precession rate by the local orbital period $2\pi r/v$. Thus we have
$\alpha_{\script{per-lap}}={\Omega}_{\rm eff} 2 \pi
r/v$. If we use this result together with \eq{hatt}, we obtain for counterclockwise
motion the clockwise angle of precession per lap
\begin{equation}\label{hatt2}
\frac{\alpha_{\script{per-lap}}}{2\pi}=\gamma \frac{(1-3M/r)}{\sqrt{1-2M/r}}-1.
\end{equation}
In particular, for the photon radius (where geodesic photons
can move on circles) at $r=3M$, we obtain a rotation angle of $-2 \pi$ per
orbit, independently of the velocity. This result is precisely what we
would expect from the discussion in Sec.~V~B. 
Equation~\eqref{hatt2} is equivalent to Eq.~(39) of \rcite{rindper}.\supcite{minicomment}

\subsection{Geodesic circular motion}
For a free (geodesic) gyroscope in circular motion around a static black hole we have, 
according to \eq{yes2}, $v^2/R=g$ where $R=r
\sqrt{g_{rr}}$. By also using $g_{rr}$ given by \eq{tro.rr} and $g$
given by \eq{gref}, we find the $\gamma$ factor for
free circular motion:
\begin{equation}\label{gam}
\gamma=\frac{\sqrt{1-2M/r}}{\sqrt{1-3M/r}}.
\end{equation}
Note that $\gamma$ becomes infinite for $r=3M$ as it should. 
If we use \eq{gam} in \eq{hatt2}, we obtain
\begin{equation}\label{hatt3}
\frac{\alpha_{\script{per-lap}}}{2\pi}=\sqrt{1-3M/r}-1.
\end{equation}
Equation~\eqref {hatt3} is an exact expression for the net precession angle per full orbit for an ideal
gyroscope in free circular motion around a static black hole. If we assume
the gyroscope to be freely ``floating'' within a satellite, analogous to
the discussion of Sec.~II~G, 
\eq{hatt3} gives the rotation
relative to a star-calibrated reference system of the satellite.
Equation~\eqref{hatt3} matches Eq.~(37) of \rcite{rindper}.

\section{Relation to other work}\label{relwork}
The standard approach to calculating gyroscope precession in special and
general relativity is to solve the Fermi equation for the spin
four-vector of the gyroscope. Even for simple applications in special relativity, such as circular
motion, the resulting equations can, however, be quite
complicated (see \rapp{Fermi}). 
In general relativity, the classical approaches to gyroscope
precession are based on approximations that assume ``weak'' gravity
and small velocities (see e.g Refs.~\onlinecite{gravitation2} and
\onlinecite{weinberg}).
The derived formalisms can therefore not be
applied accurately to, for example, a gyroscope orbiting close to
a black hole. 
Other approaches, such as that in \rcite{rindper}, are exact but specific to
circular motion. 
The approach of this paper, which is exact (assuming an ideal
gyroscope) and applies to arbitrary motion relative to a static
reference frame, is strongly linked to the more formal approaches 
in Refs.~\onlinecite{rickardfermi} and
\onlinecite{jantzen}.

\appendix
\renewcommand{\theequation}{A\arabic{equation}}       
\setcounter{equation}{0}

\section{The Fermi approach to circular motion}\label{Fermi}
In special and general relativity the spin of a gyroscope is
represented by a four-vector $S^\mu$. The Fermi transport law for
$S^\mu$ is given by
\begin{equation}\label{rall}
\frac{D S^\mu}{D\tau}= u^\mu \frac{D u^\alpha}{D\tau} S_\alpha.
\end{equation}
Here $u^\mu$ is the four-velocity of the gyroscope.
As a special relativistic application we consider motion with fixed speed $v$ along a circle in the $xy$-plane with an
angular frequency $\omega$. We assume that the spatial part of the spin
vector $(S^x,S^y,S^z)$ is in the
$xy$ plane (so $S^z=0$) and let the gyroscope start at $t=0$ on the positive
$x$-axis. Solving the Fermi equation is then (effectively) reduced to 
solving two coupled differential equations (see Ref.~\onlinecite{rickardfermi}):
\begin{subequations}
\begin{align}\label{rall3}
\frac{d S^x}{dt} &= \gamma^2 v^2 \omega \sin(\omega t)
(S^x\cos(\omega t) +S^y \sin(\omega t)) \\
\frac{d S^y}{dt} &= -\gamma^2 v^2 \omega \cos(\omega t)
(S^x\cos(\omega t)+S^y \sin(\omega t)). \label{rall4}
\end{align}
\end{subequations}
For the initial conditions
$(S^x,S^y)=(S,0)$ the solutions can
be written as\supcite{gravitation}
\begin{subequations}
\label{hmall}
\begin{align}
&\hspace{-2mm}     
S^x \nop = \nop S\left[\cos((\gamma \nop-\nop 1) \omega t)     
\nop+\nop (\gamma \nop-\nop 1)\sin(\omega \gamma t) \sin(\omega t) \right] \label{hm1}\\[2mm]
&\hspace{-2mm} 
S^y \nop=\nop S\left[\sin( (1\nop-\nop \gamma) \omega t )  
\nop-\nop (\gamma \nop-\nop 1)\sin(\omega \gamma t)\cos(\omega t) \right]. \label{hm2}
\end{align}
\end{subequations}
The first term on the right-hand side of Eq.~\eqref{hm1} and
Eq.~\eqref{hm2} respectively, corresponds to a rotation about the $z$-axis, but
there is also another superimposed rotation with a time dependent
amplitude. To find this solution directly from the coupled
Fermi equations seems rather difficult, even for this very symmetric and simple
scenario.

\renewcommand{\theequation}{B\arabic{equation}}     
\setcounter{equation}{0}                            

\section{Curvature and acceleration}\label{curveref}
Suppose that we have an upward accelerating 
reference frame. A test particle moves with velocity ${\bf v}$ along a
path, fixed to the reference frame, with the local curvature $R$ and
curvature direction $\nhat $.
We would like to express the part of the test particle's acceleration
that is perpendicular to the particle's momentary direction of motion,
relative to an inertial system in which the reference frame is
momentarily at rest.
For this purpose, we consider how the test particle will
deviate from a straight line fixed to the inertial system 
and directed in the momentary
direction of motion of the test particle.

\begin{figure}[t]
\begin{center}
\epsfig{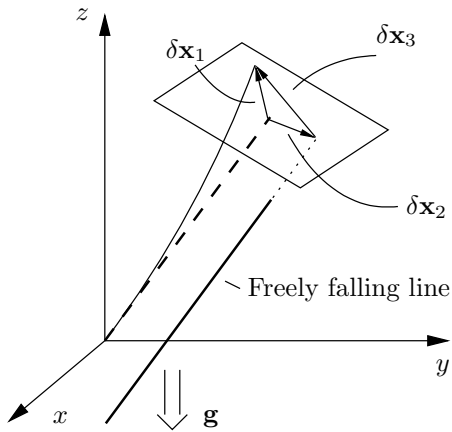}
\caption{Deviations from a
straight line relative to a reference frame that accelerates
in the $z$-direction.
The plane shown is perpendicular to the momentary direction
of motion (along the dashed line), and all the three vectors lie in this
plane. The solid curving line is the
particle trajectory.
The thick line is the line that is fixed to the inertial system in question, and is
thus falling relative to the reference frame.} 
\label{fig15}
\end{center} 
\end{figure}

For the small relative velocities between the inertial
system and the reference frame that we will consider here, we need not differentiate between the
length and time scales of the two systems.
Consider a short time step $\delta t$ after the particle has
passed the origin.
To lowest order with respect to $\delta t$, the perpendicular
acceleration relative to the reference frame is given by $v^2 \nhat/R$.
From \fig{fig15} we have 
to lowest nonzero order in $\delta t$

\begin{subequations}
\begin{align}
\delta {\bf x}_1&= \frac{\nhat}{R} \frac{v^2 \delta t^2}{2} \\
\delta {\bf x}_2&= {\bf g}_\perp \frac{\delta t^2}{2} \\
\delta {\bf x}_3&= \delta {\bf x}_1 - \delta{\bf x}_2. \label{extral}
\end{align}
\end{subequations}
Here ${\bf g}_\perp$ is
the acceleration of the inertial system relative to the reference frame (we have ${\bf
g}_\perp =-[{\bf a}_{\script{ref}}]_\perp$) in the direction
perpendicular to the direction of motion. 
We know that $\delta {\bf x}_3=[{\bf a}_{\script{particle}}]_\perp \frac{\delta
t^2}{2}$ to lowest order in $\delta t$. If we substitute this expression for
$\delta {\bf x}_3$ into Eq.~\eqref{extral} and take the infinitesimal limit,
it follows that
\begin{equation}\label{hepp}
[{\bf a}_{\script{particle}}]_\perp=[{\bf a}_{\script{ref}}]_\perp + v^2 \frac{\nhat}{R}.
\end{equation}
Equation~\eqref{hepp} gives the acceleration ${\bf
 a}_{\script{particle}}$ of a test particle, 
relative to an inertial system 
in which the reference frame is momentarily at rest,
for given path curvature relative to the
reference frame and given acceleration ${\bf a}_{\script{ref}}$ of
the reference frame.

\clearpage


\begin{thebibliography}{999}
\bibitem{actually} This angular velocity
describes how the gyroscope axis precesses if we
imagine that the length contraction along the
direction of motion is removed. We will discuss this point further in the text.
\bibitem{photrad}There is a radius where free photons, that is, photons
whose motion is determined by nothing but gravity, can move on
circles around a black hole. The circumference of this circle is $1.5$ times the circumference of the surface of the black
hole (the event horizon).
\bibitem{note}
Consider an infinitesimal boost in a general direction between the $x$ and
$y$ axis of $S'$. Because relativistic rotation effects vanish at
low speeds (according to \eq{ahhh} $\delta
\alpha \simeq v \delta u/2$), the boost is, to first order in the velocity change, equivalent to
first boosting the grid in the $x$-direction and then in the
$y$-direction. The forward $x$-boost has no impact on the relative
rotations. It follows that, to first order in the velocity change,
the $y$-part of the boost contributes to the rotation
as if there were no simultaneous forward $x$-boost.
\end{thebibliography}
\end{document}


\twocolumn[\large {}]                             
\noindent
{\bf \Large Addendum to ``Gyroscope precession in special and general relativity from basic
principles''}\\[3mm] 
{\bf Rickard M. Jonsson}\\[2mm] 
{\it Department of Theoretical Physics, Physics and Engineering
Physics, Chalmers University of Technology, and G\"oteborg University, 412
96 Gothenburg, Sweden\\[2mm] 
}
E-mail: rico@fy.chalmers.se\\[2mm]
\\
{\bf Abstract}.
I extend the reasoning of Rickard Jonsson, Am. Journ. Phys {\bf 75} 463, (2007) to include also rotating reference frames. 

\section{Rotating reference frame}
\begin{figure}[ht]
  \begin{center}
    \psfrag{S}{$S$}
    \psfrag{w}{$\omega$}
    \psfrag{a}{$a$}
    \psfrag{B}{$B$}
	\epsfig{figure=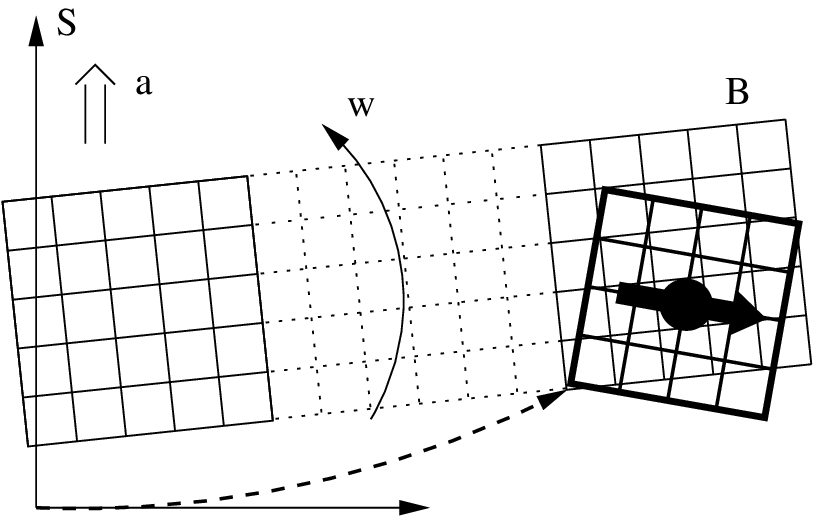,width=6cm,angle=0}
	\caption{A gyroscope grid (thick lines) moving relative to
      	a rotating and accelerating reference frame (thin lines).
	Notice that due to the rotation, the part of the reference frame situated
      	at the gyroscope grid at $t=\delta t$ has an extra velocity
      	apart from that coming from the 
	acceleration of the
      	reference frame. 
	This Coriolis related velocity will give an extra contribution
      	to the relative precession, analogous to the already discussed effect of
      	pure accelerations of the reference frame, apart from the obvious 
      	non-relativistic effects of the rotation.} 
     	\label{fig11}
  \end{center} 
\end{figure}

Suppose that the reference frame is accelerating with
acceleration ${\bf a}_{\script{ref}}$ and rotating (rigidly) with angular 
velocity {\fat{\omega}}, relative to an inertial frame $S$ momentarily comoving with
a certain point of the reference frame at $t=0$. At this time the
gyroscope (grid) is assumed to pass the point in question. The scenario is illustrated in \fig{fig11}.

Let us denote the part of the reference frame situated at the gyroscope grid at
$t=\delta t$ by $B$. Relative to $S$, $B$ has a small velocity already
at $t=0$. At $t=\delta t$, it has to lowest order the velocity ${\bf
a}_{\script{ref}} {\ } \delta t + {\fat{\omega}} \times {\bf v} \delta t$. 
To lowest order in $\delta t$ we may consider $B$ to
have been at rest in $S$ at $t=0$, but rotated by an angle $\omega \delta t$
around ${\bf \hat{\omega}}$. It is then accelerated to a velocity ${\bf
a}_{\script{ref}} {\ } \delta t + {\fat{\omega}} \times {\bf v} \delta t$, 
hence getting an extra turn from the
Thomas precession as seen from a system comoving with the gyroscope
grid at $t=\delta t$. With this insight one can modify
Eq. 12 of \rcite{intufermi},
to include the rotation. We just replace ${\bf
a}_{\script{ref}}$ by ${\bf a}_{\script{ref}} + {\fat{\omega}} \times
{\bf v}$ and add a $-\fat{\omega}$ term, to account for the mentioned
(non-relativistic) rotation, to find
\begin{eqnarray}\label{yesrot}
\fat{\Omega}= &&\hspace{-6mm}\frac{\gamma}{\gamma+1}\left(\left[
\gamma {\bf a}_\script{gyro} + \left( {\bf
a}_{\script{ref}} + {\fat{\omega}} \times {\bf v} \right)
\right] \times {\bf v} \right) \nonumber \\ &&\hspace{-6mm}- {\fat{\omega}} 
.
\end{eqnarray}
By ${\bf a}_\script{gyro}$ we here mean the acceleration of the
gyroscope relative to
the inertial frame $S$. In the non-rotating case we had $[{\bf
a}_\script{gyro}]_\perp=v^2 \frac{{\bf \hat{n}}}{R} + [{\bf a}_{\script{ref}}]_\perp$. 
Now that we have rotation this is
modified to $[{\bf a}_\script{gyro}]_\perp=v^2 \frac{{\bf \hat{n}}}{R} + 
[{\bf a}_{\script{ref}}]_\perp+2 \fat{\omega}\times
{\bf v}$. This is a simple Coriolis effect that is easy to derive
analogous to the proof in Appendix B of \rcite{intufermi}.
Using this in \eq{yesrot} yields
\begin{eqnarray}\label{yesrot2}
\fat{\Omega}=\frac{\gamma}{\gamma+1}&&\hspace{-6mm}
\Big(
\Big[
\gamma v^2 \frac{{\bf \hat{n}}}{R} + 
 {\bf a}_{\script{ref}} (\gamma + 1) 
\nonumber \\ &&\hspace{-2mm}
+(2\gamma+1) {\fat{\omega}} \times {\bf v} 
\Big]
\times {\bf v} 
\Big)
 - {\fat{\omega}} . \qquad
\end{eqnarray}
So here is the precession rate of the gyroscope grid (in the
stopped sense) relative to an accelerating and rotating reference frame. 
Comparing \eq{yesrot} and \eq{yesrot2} with the more formally
derived results of \rcite{rickardfermi} (Eqs. 50 and 51), we find a perfect match.

To apply this formalism to for instance rotating black holes, we must
take the spatial geometry into account analogous to the case for
static black holes, see Ref. \onlinecite{rickardfermi}.